\documentclass[]{aa}
\usepackage{txfonts}
\usepackage{epsfig}
\usepackage{graphicx}
\usepackage{amssymb}	
\usepackage{natbib}
\usepackage{latexsym}
\usepackage{hyperref}
\usepackage[small,hang,bf]{caption}
\bibpunct{(}{)}{;}{}{}{,}

\usepackage{graphicx}

\usepackage{xspace}

\newcommand{\halpha}{H$\alpha$\xspace}

\newcommand{\Nh}{\rm $N_{\rm\scriptsize{H}}$\xspace}

\newcommand{\degree}{\ensuremath{^\circ}\xspace}

\def\apjl{ApJ}%
\def\apjs{ApJS}%
\def\ao{Appl.~Opt.}%
\title{The northwestern ejecta knot in SN 1006}
\author{S. Broersen\inst{1}
\and J. Vink\inst{1}
\and M. Miceli\inst{2}
\and F. Bocchino\inst{3}
\and G. Maurin\inst{4}
\and A. Decourchelle\inst{5}
}

\institute{Astronomical Institute "Anton Pannekoek", University of Amsterdam, Postbus 94249, 1090 GE Amsterdam, The Netherlands \\ \email{s.broersen@uva.nl}
\and Dipartimento di Fisica, Universit\`{a} di Palermo, Piazza del Parlamento 1, 90134 Palermo, Italy
\and INAF-Osservatorio Astronomico di Palermo, Piazza del Parlamento 1, 90134 Palermo, Italy
\and Universit\'{e} de Savoie, 27 rue Marcoz, BP 1107 73011-Chambery cedex, France
\and Laboratoire AIM, CEA-IRFU/CNRS/ Univ Paris Diderot, Servic d'Astrophysique/IRFU/DSM/CEA Saclay, Gif-sur-Yvette Cedex, France
}
\date{October 26, 2012}
\abstract{}{We want to probe the physics of fast collision-less shocks in supernova remnants. We are interested in the non-equilibration of temperatures and particle acceleration. Specifically, we aim to measure the oxygen temperature with regards to the electron temperature. In addition, we search for synchrotron emission in the northwestern thermal rim. }{This study is part of a dedicated deep observational project of SN 1006 using XMM-Newton, which provides us with currently the best resolution spectra of the bright northwestern oxygen knot. We aim to use the reflection grating spectrometer to measure the thermal broadening of the $\ion{O}{vii}$ line triplet by convolving the emission profile of the remnant with the response matrix.}{The line broadening was measured to be $\sigma_{\rm e}=2.4\pm0.3$ eV, corresponding to an oxygen temperature of $275^{+72}_{-63}$  keV. From the EPIC spectra we obtain an electron temperature of $1.35\pm0.10$ keV. The difference in temperature between the species provides further evidence of non-equilibration of temperatures in a shock.  In addition, we find evidence for a bow shock that emits X-ray synchrotron radiation, which is at odds with the general idea that due to the magnetic field orientation only in the NE and SW region X-ray synchrotron radiation should be emitted. We find an unusual H$\alpha$ and X-ray synchrotron geometry, in that the H$\alpha$ emission peaks downstream of the synchrotron emission. This may be an indication for a peculiar \halpha shock, in which the density is lower and neutral fraction are higher than in other supernova remnants, resulting in a peak in \halpha emission further downstream of the shock. }{}
\titlerunning{The northwestern ejecta knot of SN 1006.}
\authorrunning{S. Broersen et al.}
\begin{document}

\maketitle

\section{Introduction}

Supernova remnants (SNRs) have the highest velocity, collision-less shocks that can be studied in the Galaxy. These shocks are of interest for the physical processes that are connected to them, including non-equilibrium effects and cosmic-ray acceleration \citep[see e.g.][]{vinkreview}. The advent of high resolution X-ray observatories such as XMM-Newton and Chandra gives us a plethora of high quality data which allow these studies to be conducted.
\begin{figure}[!h]
\centering
\resizebox{\hsize}{!}{\includegraphics[]{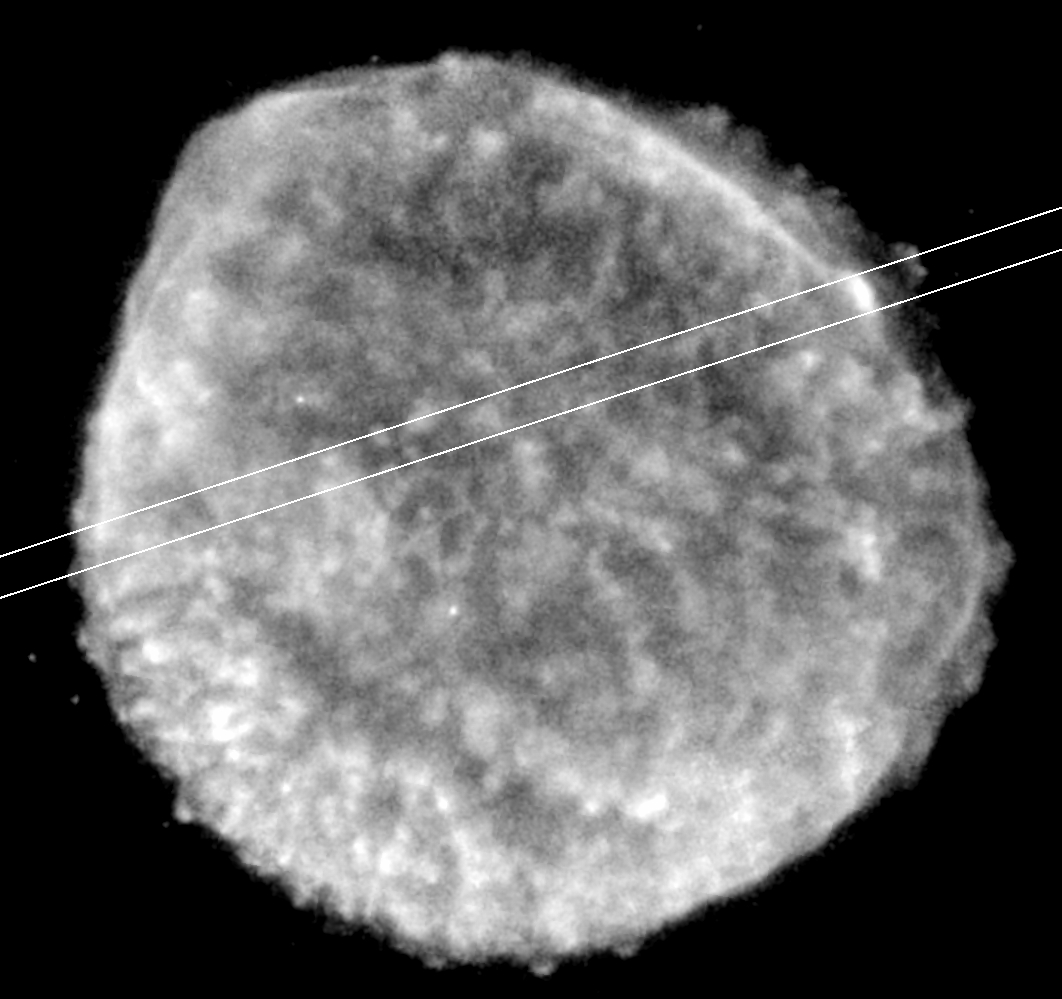}}
\caption{A smoothed image of SN 1006 in the 500-599 eV band. The bright knot in the northwest is clearly visible. The extraction region of the RGS spectrum of the 2008 observation is shown. }
\label{fig:1006}
\end{figure}

The remnant of the historical supernova  1006 A.D. (SN 1006, Fig. \ref{fig:1006}) is one of the youngest galactic SNRs in terms of its evolution. Due to its high latitude (+14.6), and its long historical lightcurve, it is thought to be a type Ia SNR. Located at a distance d = 2.2 kpc \citep{winkleretal2003}, which will be adopted throughout this paper, it is expanding in a low density local medium \citep[$\approx$ 0.15 cm$^{-1}$ in the northern region, $\approx0.05$ cm$^{-3}$ elsewhere, e.g.][]{aceroetal2007, raymondetal2007, micelietal2012}. Because of the low surrounding ISM density, SN 1006 is in an early evolutionary state and so an ideal remnant to study non-equilibrium effects of the temperature and ionisation state of the plasma. The remnant has a particular morphology, in the sense that X-ray synchrotron radiation seems to be emitted only in the northeastern and southwestern part of the remnant, with very little synchrotron emission along the line of sight toward the centre. The most viable explanation is that  these synchrotron limbs are polar caps of the remnant, and the axis of the ambient magnetic field lies SW--NE \citep[][]{rothenflugetal2004, volketal2003, Berezhkoetal2009, bocchinoetal2011}. Such a magnetic field parallel to the shock makes injection and thus acceleration of particles more efficient, creating a higher density of accelerated particles at the poles \citep{ellisonetal1995}. In addition, the $\gamma$--ray emission shows the same morphology \citep{1006_HESS_2010}. 

This study is part of a large observing project on SN 1006 (PI: dr. A. Decourchelle), which consists of 7 XMM Newton pointings which, coupled with archival data, bring the total observing time to $\approx900$ ks. Here we focus on the bright emission knot in the northwestern part of SN 1006. This interesting region of the remnant has been studied before in detail by \cite{longetal2003} with Chandra. The knot itself has been studied  in more detail by \cite{vinketal2003}, who used the Reflection Grating Spectrometer (RGS, \cite{denherderetal2001}) to determine the ion temperature based on the thermal broadening of the $\ion{O}{vii}$ line triplet. Since we now have a factor four more observation time, we are able to study the knot in much more detail and with higher precision. Since the knot is also observed with different roll angles we have a better control of systematic effects. 

In their study of the emission knot, \cite{vinketal2003} found $T_{\rm Oxygen} = 530 \pm 150$ keV, while the electron temperature $T_{\rm e}$ was measured at $1.5\pm0.2$ keV. A temperature as high as $530$ keV suggests that the gas has been shocked with a velocity in excess of $\approx5000$ km s$^{-1}$, provided that all the shock energy goes into heating the plasma. 
The result of \cite{vinketal2003} confirmed earlier results in the optical  \citep{ghavamianetal2002} and UV \citep{raymondetal1995}, namely that particle species in shocks are heated proportional to their mass i.e. their temperatures are not equilibrated:
\begin{equation}
\label{for:1}
kT_{\rm i} = \frac{2(\gamma-1)}{(\gamma+1)^2}m_{\rm i}v_{\rm s}^2 = \frac{3}{16}m_{\rm i}v_{\rm s}^2,
\end{equation}
where k is Boltzmann's constant, $T_i$ is the specie's temperature, $\gamma$ is the equation of state of the plasma (5/3 for non relativistic matter), $v_s$ is the shock velocity and $m_i$ is the specie mass. In addition, the result shows that the equilibration of temperatures behind the shock front in SNRs is a slow process. 

In addition to the ejecta knot, we report on the X-ray emission upstream of the knot which shows evidence for X-ray synchrotron radiation. 

\section{Data analysis}
\label{sec:data_analysis}
For studying the NW knot we use the three pointings done in 2001, 2008 and 2009, with OBSID's 0077340101, 0555630501 and 055530401 respectively. Both the EPIC (MOS and pn) and the reflection grating spectrometer (RGS) where used for this study. 

The RGS1 and 2 data were scanned for soft-proton flaring using CCD nr.\ 9 of the detector, which is closest to the optical axis of the mirror and thus most sensitive to flaring. The total, reduced, observation time of the RGS data amounts to 220 ks. For the knot spectrum, we extracted only the part of the CCD's where the knot emission is present, based on an image in the 500-600 eV band.

The large extent of SNR 1006 creates difficulties in the data analysis in two ways. Firstly, the background is normally determined by taking an extraction region from the edges of the CCD's. This is impossible for SN 1006, because the edges of the CCD also contain emission from the remnant. Since the source is much brighter than the cosmic background, we take a - flaring corrected - blank sky observation (OBSID 0500630101, 74.2 ks observation time) from a similar orbit  to account for the background. 

The second problem is that the RGS is a slitless spectrometer. As a result, the lines get smeared out due to the large angular extent of the source. The source diameter of 30' results in a line broadening of $\approx$4 \AA. This emission is in addition to the bright emission of the knot, which has an extent of 0.45'. We used an updated version of the code used in \cite{vinketal2003}, to correct for the source extent. This code convolves the response matrix with the emission profile of the remnant in the direction of the RGS dispersion axis, taking into account the vignetting and off-axis efficiency of the instrument. The emission profile was obtained using a EPIC mosaic. While doing the data analysis, we noticed that the off-axis emission was not symmetric between the observations, which have a roll angle difference of $180\degree$. We corrected for this based on the RGS vignetting calibrations documented in tn-cal-98\_002 calibration document \footnote{Available at http://www.sron.nl/divisions/hea/xmm/internal/docu-ments/rgs-sron-tn-cal-98\_002.pdf}.

For the data analysis, we used the SPEX spectral fitting code \citep{spex}.

\section{Results}
\label{sec:results}

\begin{figure}
\centering
\resizebox{\hsize}{!}{\includegraphics[angle=0]{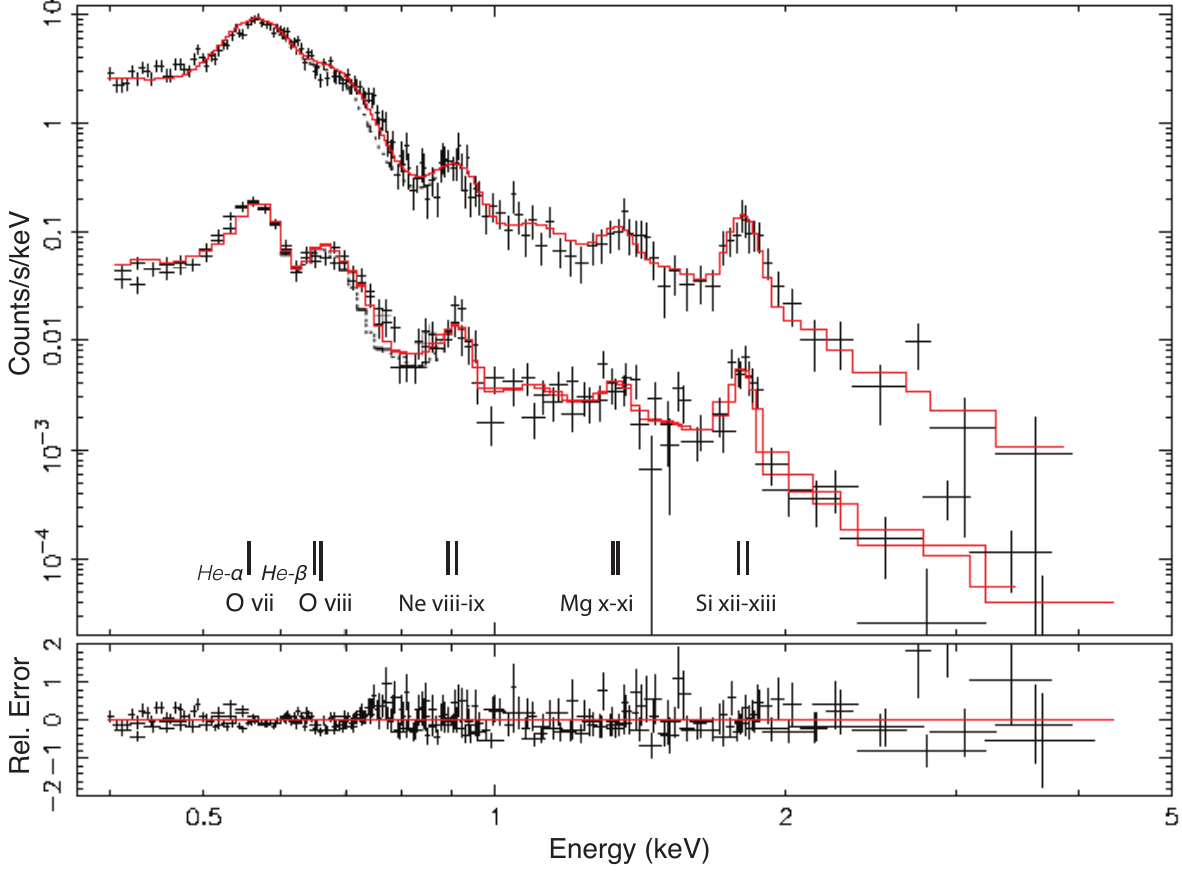}}

\caption{The EPIC MOS and EPIC PN ($ \mbox{x}10$) spectrum of the knot in SNR 1006. For clarity we only show the spectra of the 0555630401 observation. The dotted line shows the model without the higher order $\ion{O}{vii}$ lines (see section \ref{sec:results}).}
\label{fig:epic_spec}
\end{figure}

\subsection{Overall spectra}
We extracted a MOS and pn spectrum (Fig. \ref{fig:epic_spec}) from the knot for the three different epochs. The spectrum is dominated by emission lines of the lower mass elements such as $\ion{O}{vii/viii}$ and $\ion{Ne}{viii/ix}$, although there is a significant contribution from Li- and He-like $\ion{Mg}{}$ and $\ion{Si}{}$. Because of the low surrounding density, the ionisation age of SN 1006 is one of the lowest measured in the Galactic supernova remnants. We used C-statistic to fit the data, which for a progressively higher number of counts asymptotically approaches the $\chi^2$ value \citep{cash}.  The best fit model to the EPIC spectrum of the knot is a single Non-Equilibrium Ionisation (NEI) model and a basic absorption model. The best fit \Nh we find is comparable to the values found by \citep{dubneretal2002}.

%with the hydrogen column \Nh fixed at $6.8\times10^{20}$ cm$^{-2}$ \citep{dubneretal2002}. 

Because the knot is assumed to be an ejecta knot, it is not expected that more than one temperature NEI component is present. The best fit parameters are listed in Table \ref{tab:abundances}. The ionisation age of $3.1\times10^9$cm$^{-3}$ s and temperature of 1.35 keV are compatible with the values found by \cite{vinketal2003}. 

\begin{figure}[!t]
\centering

\resizebox{\hsize}{!}{\includegraphics[]{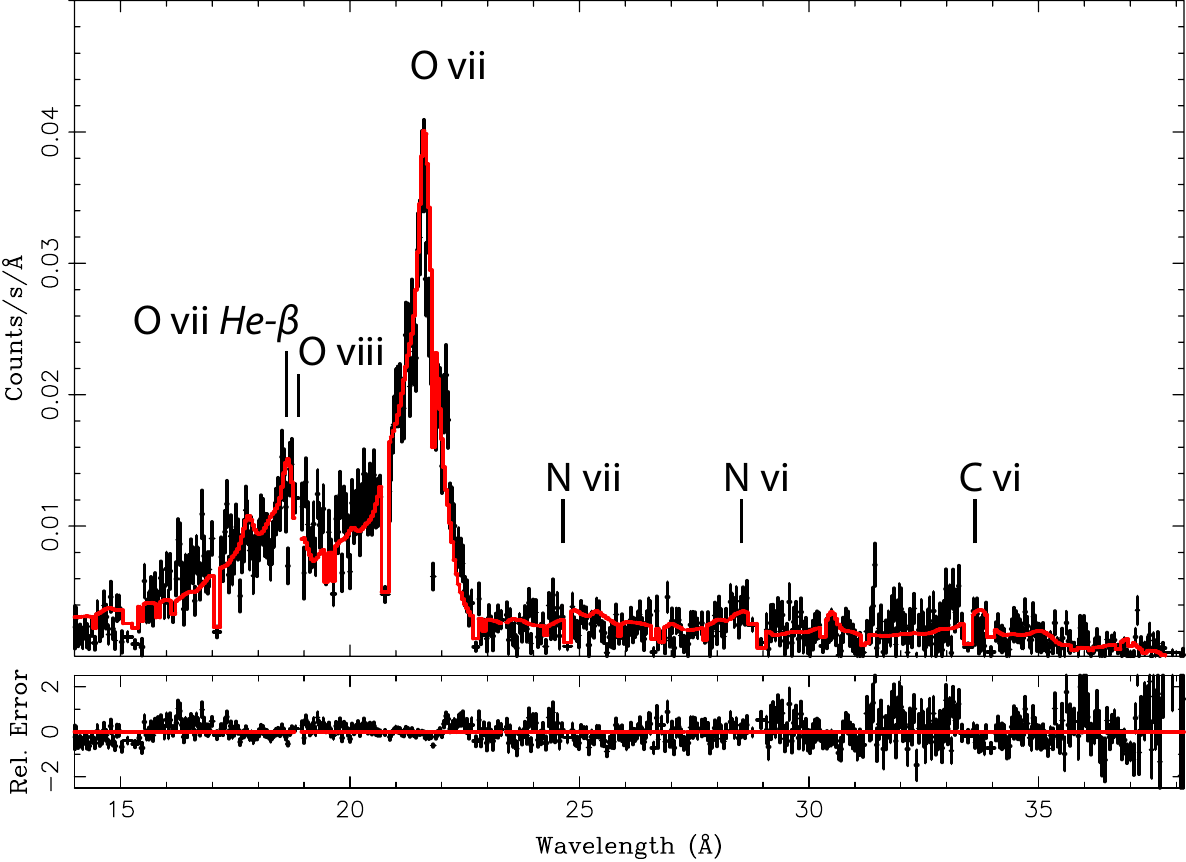}}
\caption{RGS spectrum of the 2008 observation fit by a single NEI model.}
\label{fig:0401_RGS_spec}
\end{figure}

\begin{figure}[h!]
\centering
\resizebox{\hsize}{!}{\includegraphics[]{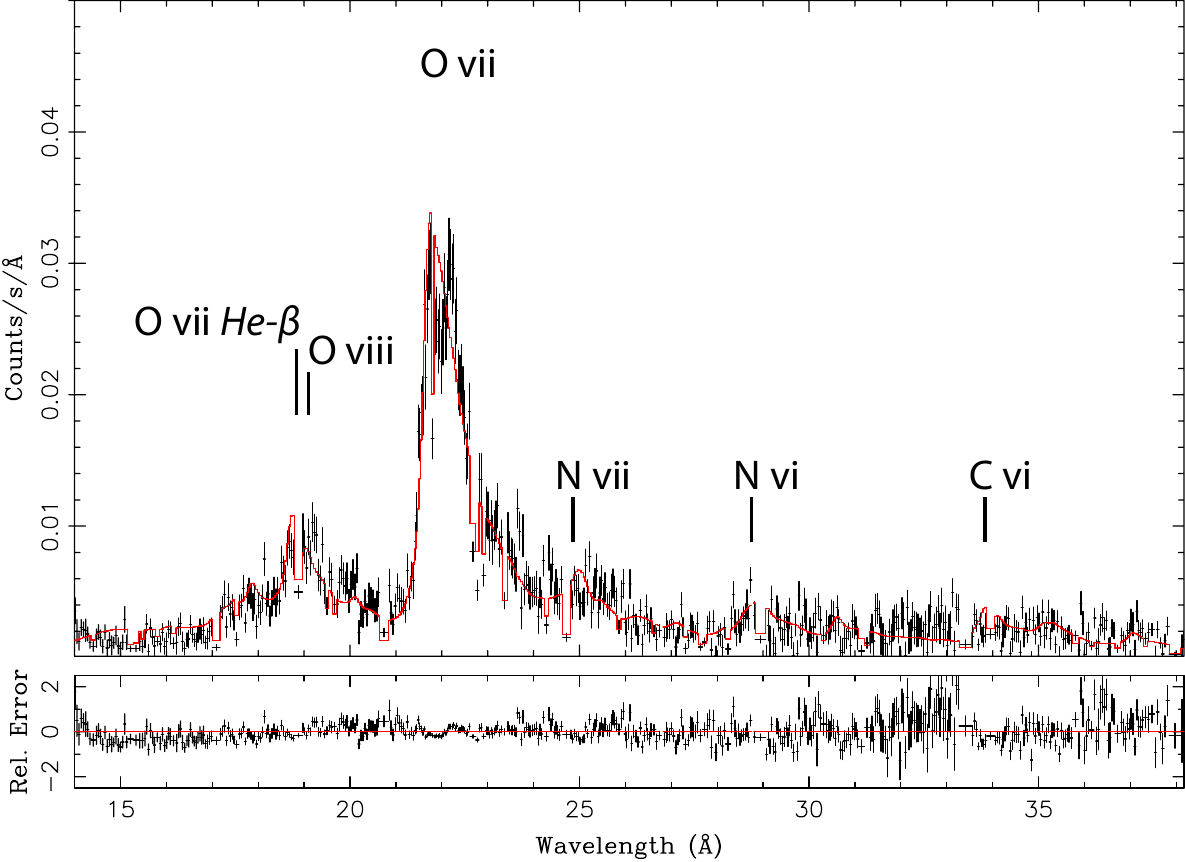}}
\caption{RGS spectrum of the 2009 observation fit by a single NEI model.}
\label{fig:0501_RGS_spec}
\end{figure}

The RGS spectra of the oxygen knot are shown in figures \ref{fig:0401_RGS_spec} and \ref{fig:0501_RGS_spec}. The difference in the spectra due to the 180\degree difference in roll-angle is clearly visible; the excess emission which results from the extent of the remnant shows on different sides of the $\ion{O}{vii}$ line triplet. The same is true for the emission around the other $\ion{O}{}$lines at 19 and 18.6 \AA. As in the EPIC spectra, $\ion{O}{}$is clearly the dominant element in terms of emission in the RGS spectra. No other elements have such prominent emission lines, although there is some $\ion{Ne}{IX}$ present at 13.6 \AA~and there are hints of emission lines in the higher wavelength part of the spectrum, above $23~\AA$. Although the correction described in section \ref{sec:data_analysis} works well for the emission lines, there may still be continuum emission from the synchrotron bright part of the remnant which cannot be disentangled with our method. This results in a systematically higher continuum emission in the RGS spectrum, which in turn results in lower abundances found by fitting the RGS data. We therefore list only the values found by fitting the EPIC data.  The ionisation age and electron temperature found are consistent with the MOS data. 

In previous studies of the spectrum of SN 1006, excess emission (see Fig. \ref{fig:epic_spec}) has been found in the 0.73-0.8 keV energy range  \citep[e.g.][]{yamaguchietal2008, micelietal2009}. It has been interpreted as both $\ion{Fe}{}$ emission and as higher order $\ion{O}{vii}$ transitions missing in the plasma code. At 0.739 keV also a recombination edge of $\ion{O}{vii}$ can be found. However, since the remnant is in an ionizing state, significant recombination is not to be expected. 

\begin{table}
\renewcommand{\tabcolsep}{.2cm}
\renewcommand{\arraystretch}{1.2}
   \begin{centering}
   \caption{\small Best fit parameters for the NEI model. }
 	\begin{tabular}{l@{ }ll}

\multicolumn{2}{l}{Parameter} & EPIC \\
\hline
\Nh &$(10^{20}$ cm$^{-2})$& $4.16\pm0.77$  \\
$n_{\rm e}n_{\rm h}V$& $(10^{54}$ cm$^{-3})$ & $2.40\pm0.18$\\
$n_{\rm e}t$& $(10^{9}$ cm$^{-3}$ s) & $3.10\pm0.07$\\
$kT$& (keV) &$1.35\pm0.10$\\
\multicolumn{2}{l}{C}&$0.29\pm 0.08$\\
\multicolumn{2}{l}{N}&$0.57\pm0.06$ \\
\multicolumn{2}{l}{O}&$0.69\pm0.04$\\
\multicolumn{2}{l}{Ne}&$0.12\pm0.01$\\
\multicolumn{2}{l}{Mg}&$0.45\pm0.05$\\
\multicolumn{2}{l}{Si}&$4.75\pm0.37$\\
%\multicolumn{2}{l}{Ca}&$2.33\pm0.21$\\
%\multicolumn{2}{l}{Fe}&$0.99\pm0.17$\\
\hline
\multicolumn{2}{l}{cstat / d.o.f.}& 1603 / 842  \\
\end{tabular}
    \tablefoot{Best fit parameters for the EPIC data. The MOS and pn data of the three observations of the knot were fitted simultaneously. The \cite{abundances} Solar abundances were used and the errors denote a $1\sigma$ uncertainty. Non listed abundances were fixed at Solar. }
    %The \Nh was taken from \cite{dubneretal2002}. }	\\
        
    \label{tab:abundances}
\end{centering}
\end{table}

Of the listed possibilities, the Fe explanation seems the least plausible. Especially in the ejecta knot, the ionisation age is so low that Fe has not been ionised enough to emit significantly in the X-ray part of the spectrum (i.e. its ionisation state is below $\ion{Fe}{xvii}$). In addition,  An $\ion{Fe}{xvii}$ emission line at this energy would surely be accompanied by other $\ion{Fe}{xvii}$ emission lines \citep{gillaspyetal2011}, for which we do not find evidence. 

The missing $\ion{O}{vii}$ transitions seem to be the most probable explanation to solve the excess. Higher order K-shell transitions (He-$\epsilon$, He-$\zeta$ and up) are indeed missing in the current version of SPEX. Fitting the spectrum with a preliminary version of SPEX in which the new lines are included significantly improved the fit. As opposed to the gaussians added by \cite{yamaguchietal2008}, the ratio between the lines, i.e. He-$\zeta/$He-$\epsilon$, in a plasma of this temperature and ionisation age, may be as high as 0.75. The ionisation age and kT of the spectral fit do not change significantly between the older and the newer version. As the new version of SPEX is still preliminary, we list in Table \ref{tab:abundances} the parameters obtained by fitting with the current public version of SPEX. 

Ejecta knots are expected to have large overabundances of elements. Looking at the abundances, the expected ejecta products $\ion{Si}{}$ and $\ion{Ca}{}$ are found to have higher than solar abundances. The other elements, most notable $\ion{Fe}{}$ have abundances below or at solar values. Both the MOS (at 4.4$\sigma$) and the RGS (at $>10\sigma$) data show statistical evidence for the presence of $\ion{N}{}$. The presence of this element can be related to interaction with ISM, as it is not created in type Ia SNe. 

\begin{figure}
\centering
\resizebox{\hsize}{!}{\includegraphics[]{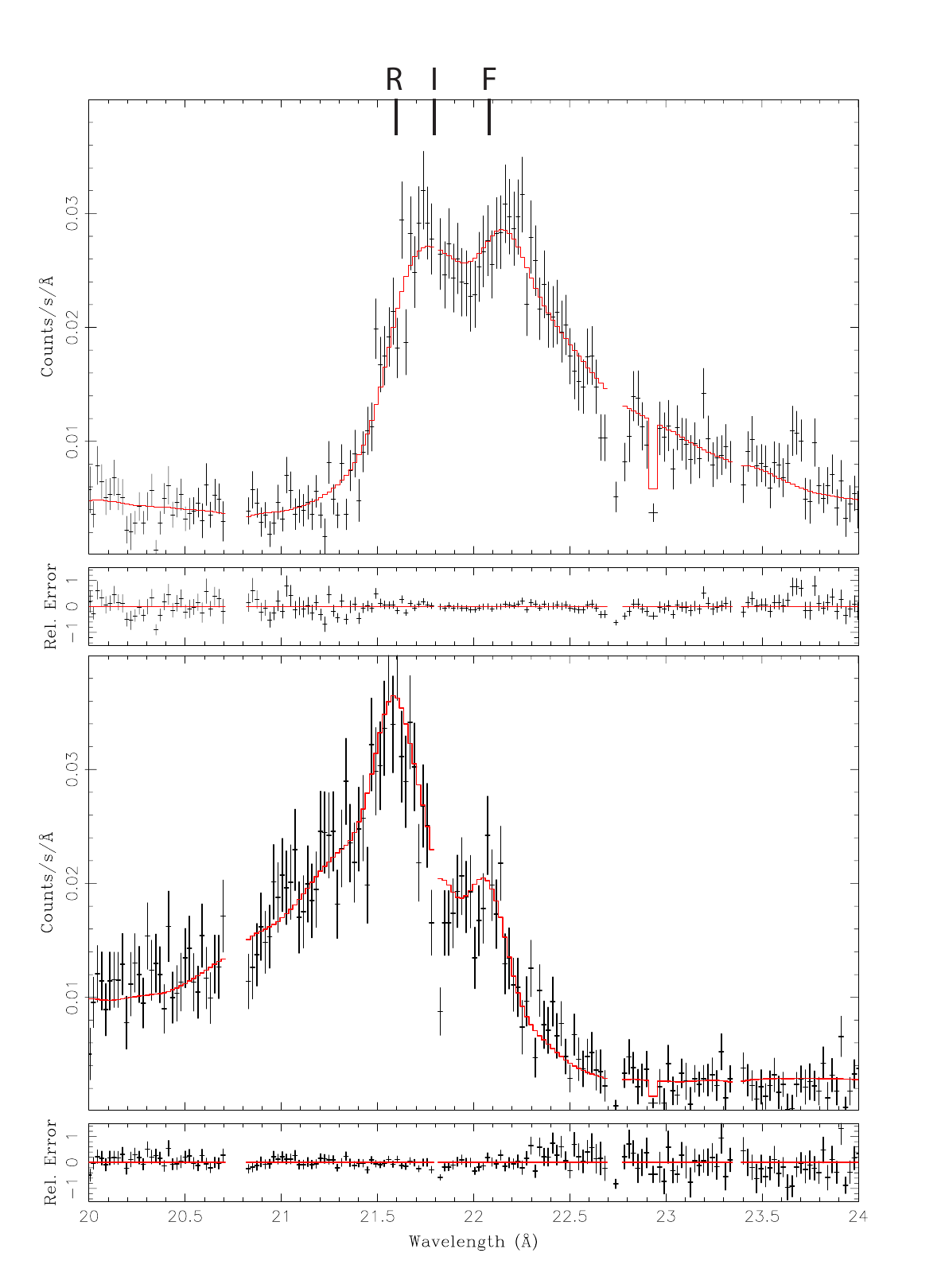}}
\caption{Zoom of the $\ion{O}{vii}$ line triplets of the 2008 and 2009 observations. The model fitted to the data is identical. The apparent difference in line ratios and model is due to the 180\degree difference in roll-angles between the two observations. The positions of the Resonance (21.6 \AA), the Intercombination (21.8 \AA) and the Forbidden line (22.1 \AA) are indicated in the figure. }
\label{fig:triplet}
\end{figure}

 \subsection{Ion temperature}
Fig. \ref{fig:triplet} shows a close-up of the $\ion{O}{vii}$ line triplet for the 2008 and 2009 observations. The triplet was fit with three Gaussians and the best fit continuum from the complete dataset fit. This gives a best fit doppler broadening of $\sigma_{\ion{O}{vii}}=2.4\pm0.3$ eV for the 2008 and 2009 observations combined, which corresponds to a $kT_{\ion{O}{vii}} = 275^{+72}_{-63}$ keV \footnote{Obtained with the formula $\sigma_{\ion{O}{vii}}/E_{0} = \sqrt{(kT/mc^2)}$, with $E_0$ = 574 eV and $m = m_{\rm Oxygen} \approxeq 16m_{\rm p}$.}. This is substantially larger than measured the electron temperature of 1.2 keV. The line broadening was significant at a $>7\sigma$ level. The new value of $kT_{\ion{O}{vii}} = 275^{+72}_{-63}$ is smaller than the value reported by \cite{vinketal2003}, but the measurements are consistent at the $1.5\sigma$ level. Our new measurement is more reliable, as not only the statistics have improved, but also, because the observations were taken with different roll angles, the systematic errors are reduced. As we experimented with different extractions, and different model assumptions (a NEI model, or a model in which line emission was modeled with gaussians in which both intensity ratios and line widths were not fixed), we were able to estimate that the systematic error is $\pm45$ keV.

\subsection{Bowshock}
\label{sec:bowshock}
\begin{figure}
\centering
\resizebox{\hsize}{!}{\includegraphics[]{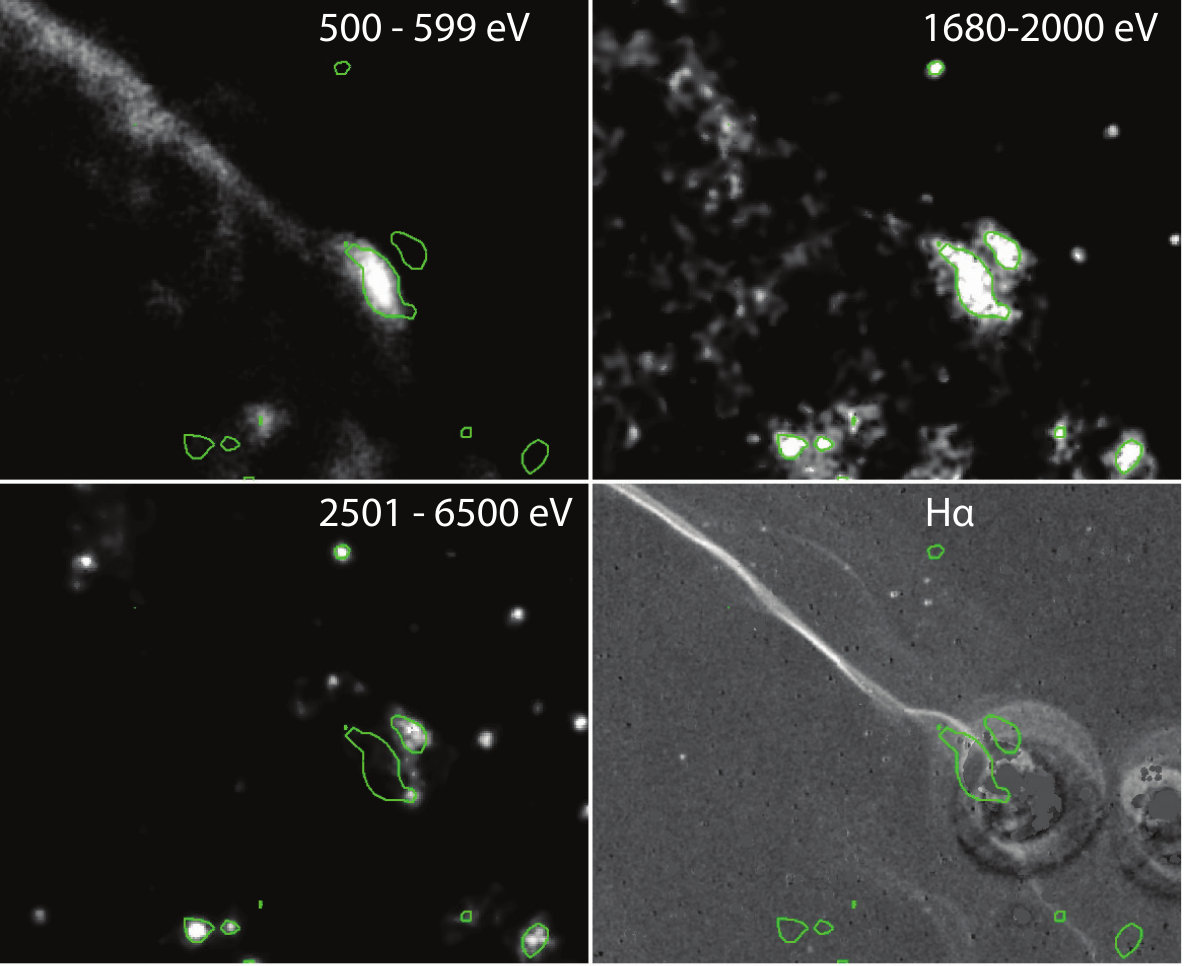}}
\caption{The emission knot plus the hard X-ray region in front of it seen in the $\ion{O}{vii}$ (500-599 eV), $\ion{Si}{xii-xiii}$ (1680-2000 eV), continuum (2501-6500 eV) and H$\alpha$ wavebands. The 1680-2000 eV band contours are plotted in all regions for comparison. The ring like structure in the H$-\alpha$ image is caused by the removal of a foreground source.}
\label{fig:bowshock}
\end{figure}

\begin{figure}
\centering
\resizebox{\hsize}{!}{\includegraphics[angle=0]{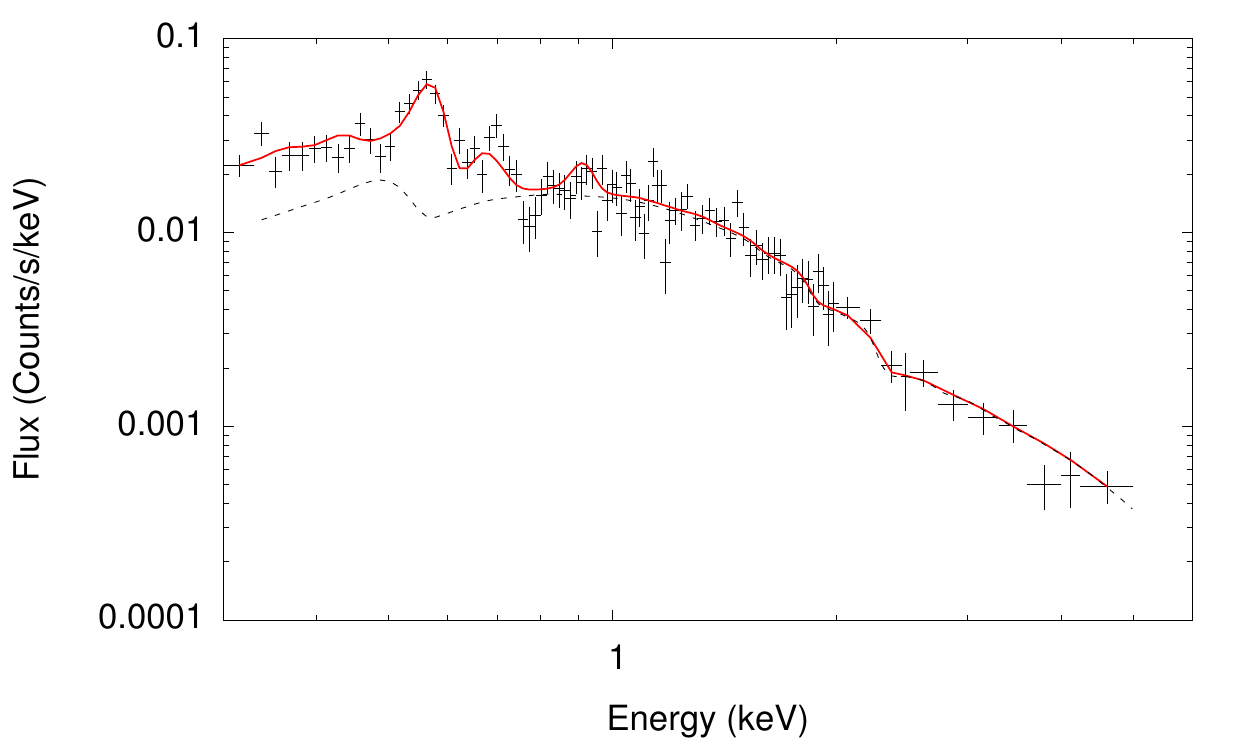}}
\caption{Total added MOS spectrum of the hard X-ray region in front of the knot. The best fit model to the spectrum includes a power law and an NEI component The powerlaw contribution is shown as a dotted line.  }
\label{fig:bs_spec}
\end{figure}

Fig. \ref{fig:bowshock} shows from softer to harder X-ray bands and H-$\alpha$ emission the region of the ejecta knot. The $\ion{O}{vii}$ and H-$\alpha$ bands clearly show a bow-shock like protrusion in the shock front, affiliated with the ejecta knot. The 1680-2000 eV band supports the picture, formed by the abundances, of an almost pure Si ejecta knot overtaking the shock front. An interesting feature is present upstream of the H$\alpha$ filament: an isolated patch of hard X-ray emission which is not visible in the $\ion{O}{}$band. A MOS spectrum of the contour region which lies exactly on the bright spot upstream of the knot in the 1680-2000 eV band, is shown in Fig. \ref{fig:bs_spec}. The spectrum shows $\ion{O}{vii-viii}$ and some $\ion{Ne}{viii}$ emission, but there is an absence of emission lines from higher mass elements, which suggests that the hard X-ray component upstream of the knot is non-thermal emission. 

Contrary to the ejecta knot, a single NEI model is not sufficient to account for the emission in the bow shock. Our best fit model contains a power law component, with a spectral index $\Gamma=2.34\pm0.06$, at a cstat / d.o.f. = 141/131. A powerlaw fit is preferable over a hot NEI component with a significance of $>5\sigma$. The NEI model has $kT = 0.80_{-0.28}^{+0.59}$ keV and $n_{\rm e}t = 1.5^{+0.7}_{-0.3}\times10^9$ cm$^{-3}$. The abundances were fixed at solar. The temperature and ionisation age of the NEI component are lower than those found in the bright emission knot, and are somewhat similar to the ISM spectral component by \cite{micelietal2012}. The spectrum shown in Fig. \ref{fig:bs_spec} has a similar background region as the spectrum of the ejecta bullet. To check whether the lower energy emission lines in the bow-shock spectrum are due to scattered light, however, we also considered a background spectrum taken from a region located at the same distance from the bullet as the bow shock. Indeed the lines disappeared in this case, while the powerlaw remained. Since Fig. \ref{fig:1006} clearly shows diffuse $\ion{O}{}$ emission around the ejecta knot, but also upstream the northwestern H$\alpha$ shock front in general, it remains unclear whether the thermal emission in the spectrum can be fully attributed to scattered light, or if part of the emission is indeed due to some low temperature plasma component. For the estimates of the non-thermal flux the details of the background subtraction does not make much difference, as the fit to the power-law component was identical in both background cases. In the next section we take a more detailed look at the implications of the presence of the non-thermal emission in a bow shock upstream of the ejecta bullet.

\section{Discussion}

We performed a detailed spectral analysis of both the EPIC and RGS data of the bright northwestern knot in SN 1006. We found statistical evidence for ISM interaction due to the presence of $\ion{N}{}$emission. We find that the excess found around 0.73 keV is very likely caused by missing emission lines in the NEI codes. We measured the line broadening at the northwestern bright emission knot of SN 1006 and did a detailed spectral analysis of both the knot and the hard patch of X-ray emission in front of it. The line broadening was measured at $\sigma_{\ion{O}{vii}}=2.46\pm0.3$ eV, which corresponds to an ion temperature of $ 275^{+72}_{-63}$  keV. The electron temperature was measured to be $1.35\pm0.1$ keV, confirming that temperatures between species with different mass are not equalized. Our results are consistent with the previous temperature measurement of \cite{vinketal2003}. 

If Eq. \ref{for:1} holds (i.e. $T_{\rm p} = T_{\rm \ion{O}{}}/16$), then $T_{\rm e} / T_{\rm p} = 0.07\pm0.01$. This is similar to the value found by \cite{ghavamianetal2002}:  $T_{\rm e} / T_{\rm p} < 0.07$. However, it should be mentioned that our X-ray measurement concerns the plasma more downstream from the shock front than the optical measurement, and a different region of the northern shock front. The timescale for proton and oxygen equilibration starts to kick in at a shock age of $\approx10^{10}$ cm$^{-3}$ s, while our measured ionisation age is $3.1\times10^9$ cm$^{-3}$ s \citep[e.g.][]{vinkreview}. Electrons are, however, expected to have risen in temperature quite significantly at such a shock age, from Coulomb interactions alone. From the temperature and using Eq. 1, we can calculate the velocity at which the plasma in the knot has been shocked, $v_s = 3000$ km/s. Since it is an ejecta knot, this velocity represents the reverse shock velocity in the frame of the ejecta knot. 

We can calculate the density of the ejecta knot from the EPIC emission measure, assuming that the knot is shaped as an ellipsoid. Using the already adopted distance of 2.2 kpc and dimensions of the knot of $(4.3\times10^{17}) \times (4.3\times10^{17}) \times  (1.1\times10^{18}) $ cm$^{3}$, we obtain a volume of $\approx1.0 \times 10^{54}$ cm$^{3}$. An emission measure of $n_{\rm e}n_{\rm H} V = 3.0(\pm0.2)\times10^{54}$ cm$^{-3}$ gives a density, using $n_{\rm e}$= $n_{\rm H}/1.2$, of $n_{\rm H}  \approxeq 2$ cm$^{-3}$. This is high compared to the density of the surrounding medium, which is around 0.1 cm$^{-3}$.  An ejecta knot needs a density enhancement of $\approx100$ times with respect to the normal ejecta density at the same radius in order to survive the instabilities caused by its interaction with the reverse shock \citep{wangchevalier2001}. Magnetic field pressure, however, limits the formation of instabilities, decreasing the density contrast needed for an ejecta knot to reach the forward shock \citep{orlandoetal2012}.

\subsection{X-ray synchrotron emission}

The hard X-ray emission  found ahead of the knot is interesting in its own regard. The morphology resembles a bow shock, suggesting it may be caused by the ejecta knot, although a line of sight explanation is also possible. The parameters of the NEI model favour shocked ISM plasma as its origin. For the power law component, a different origin than synchrotron emission is not probable. Both inverse Compton and non-thermal Bremsstrahlung are not significantly present at the concerned energies \citep[see e.g. Fig. 8 in][]{1006_HESS_2010}, even if their contribution there is underestimated by two orders of magnitude. 

We can estimate the magnetic field strength at the bow shock by assuming that the width of the synchrotron patch is equal to the advection length of synchrotron emitting electrons \citep[Eq. 62 in][]{vinkreview}:
\begin{equation}
B \approxeq 26\left(\frac{l_{adv}}{1.0\times10^{18} ~\rm cm}\right)^{-2/3}\eta^{1/3}\left(\chi_4 - \frac{1}{4}\right)^{-1/3}  \rm \mu G,
\end{equation}
where $\chi_4$ is the shock compression in units of 4 and $\eta$ the deviation from Bohm diffusion, which is on the order of 1. With our observations, we can obtain an upper limit to the width of the synchrotron emitting region at $\approx10.8''$, which corresponds to an $l_{\rm adv} = 0.37\times10^{18}$ cm. Taking a shock compression ratio of 4, we obtain a lower limit to the magnetic field of $\approx50~ \mu$G. This is consistent with the magnetic field strengths measured in other parts of the remnant \citep{1006_HESS_2010}. 

Another interesting feature is, as mentioned in section \ref{sec:bowshock}, the fact that the X-ray synchrotron emission is found ahead of the H$\alpha$ emission. The H$\alpha$ emission is usually found in a narrow ($\approx10^{15}$ cm) region behind the shock front, while X-ray synchrotron emission starts behind the shock, but is much broader ($10^{17}-10^{18}$ cm). That the geometry is different near the knot could be coincidental; i.e.  due to a line of sight effect. In that case the X-ray synchrotron emission is not causally connected to the knot and the \halpha filament. It is unlikely
that the \halpha emission and the knot are not causally connected, for the \halpha filament shows a clear protrusion where the knot penetrates the shock front. 
Nevertheless, the fact that the only clear X-ray synchrotron filament not connected to the bright rims is just ahead of the ejecta knot, makes it
worthwhile to investigate whether they could be causally connected and what then the origin could be for the peculiar geometry, with
the X-ray synchrotron emission lying ahead of the \halpha filament.
 %However, such an isolated X-ray synchrotron feature right upstream of the ejecta knot would be a peculiar coincidence. We explore therefore the possibility that there is a causal connection between the X-ray synchrotron emission, the knot and the \halpha filament, and speculate on a possible cause for the particular shock geometry.}

 A causal connection between X-ray synchrotron emission and the ejecta knot (i.e. the synchrotron emission is caused by a bow shock),
may provide an explanation for why only in this region outside the bright synchrotron limbs we see evidence for X-ray synchrotron emission.
It is likely that near the knot the shock velocity is higher than in the immediate surroundings
\citep[$\approx2900$ km/s,][]{ghavamianetal2002}. Given the dependence
of X-ray synchrotron emission on the shock velocity \citep[$h\nu_{max}\propto V_{\rm S}^2$, e.g.][]{Zirakashviliaharonian2007,vinkreview}, the higher shock velocity may have resulted in a higher synchrotron cut-off frequency, and, hence,
X-ray synchrotron emission.
In addition, the interaction of the knot with its surroundings may have enhanced the
local magnetic field. For example, hydrodynamical simulations indicate that the magnetic field strength may be amplified upstream of an ejecta knot \citep{orlandoetal2012}. On the other hand, the magnetic field estimate provide above, indicates that  the magnetic field is not significantly different from the bright synchrotron rims, although the magnetic field may be smaller elsewhere along the northwestern rim.

The presence of X-ray synchrotron emission at this location is unexpected, given
the overall evidence that suggests a polar cap geometry for synchrotron 
emission; with the polar caps being the two bright X-ray synchrotron emitting
region. The general idea is that for these caps the magnetic field is parallel
to the shock normal, causing more particles to be injected into
the acceleration process. However, this model still applies 
to the emission of X-ray synchrotron radiation 
from the specific spot in the northern region. The reason is that
the X-ray emission, and also the radio emission from this region is 
relatively weak, implying that the number density of relativistic is low.
X-ray synchrotron emission itself, however, informs us that the particles
can be accelerated to 10-100 TeV energies. For this to happen the shock velocity
needs to be high enough 
\citep[$V_{\rm s}\gtrsim 3000$~km\,s$^{-1}$][]{aharonian1999,vinkreview}. 
Radio expansion measurements indicate that the 
northeastern region and southwestern region have lower expansion velocities \citep{moffetetal2004}. So a possible explanation for the lack of X-ray synchrotron 
emission from the northeastern and southwestern region is a generally low
velocity, limiting the maximum electron energy. 
A possible exception would then be the shock region coinciding with the knot. 
The overall radio and X-ray synchrotron morphology of SN 1006 is more 
determined by the efficiency of acceleration, i.e. the number density of 
relativistic  electrons, in agreement with the polar cap model \citep{rothenflugetal2004,bocchinoetal2011}.
 
A possible causal connection between the ejecta knot and the X-ray synchrotron emission, raises the question why the 
\halpha\ emission is then emitted so far downstream of the shock. Here we speculate on what could cause the peculiar observed X-ray/\halpha\  geometry.
There are several factors which determine the width of an H$\alpha$ filament \citep[e.g.][]{vanadelsbergetal2008}, which are shock velocity, neutral fraction and the pre-shock density. The width of the filament becomes larger when the velocity $v_s$ or neutral fraction $f_{\rm n}$ increase and/or when the pre-shock density n$_0$ decreases. For example, with $v_s = 4000$ km/s,  $f_{\rm n} = 0.9$ and $n_0 = 0.01$ cm$^{-3}$, the distance between the shock front and H$\alpha$ filament $z \approxeq 5\times10^{17}$ cm. This is comparable to the distance between the filament and the outer edge of the synchrotron emission $d \approxeq 8\times10^{17}$ cm (taking into account an expansion rate of the H$\alpha$ filament of 0''.3 / year). The brightest H$\alpha$ region could then correspond to the increased density associated with the contact discontinuity between the knot and the shocked ISM. This explanation does however require a high neutral fraction coupled with a low density, while the neutral fraction as measured by \cite{ghavamianetal2002} equals 0.1, and $n_0$ estimates in the northern region lie around $0.15 - 0.25$ cm$^{-3}$. On the other hand, these estimates are based on a conventional explanation for the H$\alpha$ emission, and apply to other regions in the northwestern part of SN 1006. More detailed modeling is necessary to see whether extended H$\alpha$ emission, due to a high neutral fraction, is a viable model for the H$\alpha$ emission associated with the shock, or whether the unexpected geometry of H$\alpha$ and X-ray synchrotron can be attributed to an accidental superposition of two separate features.

\section{Conclusions}

We have presented here our analysis and interpretation of a deep XMM-Newton observation of the northwestern region of SN 1006, focussed
on a bright ejecta knot. The analysis concerned both imaging spectroscopy with the EPIC instrument, and high resolution spectra of the RGS instrument.
We summarize here the main results of our analysis:
\begin{itemize}
\item The line broadening of the $\ion{O}{vii}$ ions was measured at $2.46\pm0.3$ keV, which corresponds to a $ 275^{+72}_{-63}$  keV. The electron temperature was measured at $1.35\pm0.1$ keV. Our results therefore confirm that the temperatures between species of different mass are not equilibrated.
\item We find a bow-shock of X-ray synchrotron emission upstream of the ejecta knot. X-ray synchrotron emission at this location is unexpected and at odds with the general \emph{polar cap} model for X-ray synchrotron emission \citep{rothenflugetal2004}. 
\item The shock near the ejecta knot has a peculiar geometry, in that the X-ray synchrotron emission is located upstream of a bright \halpha filament.
\item This geometry can either be explained by a line of sight effect, where the knot and the synchrotron emission are not causally connected or, more speculatively, by \halpha\ emission from a region with a high interstellar medium neutral fraction and  a low density. 
\end{itemize}

\begin{acknowledgements}

S.B. is supported financially by NWO, the Netherlands Organisation for Scientific Research. The results presented are based on observations obtained with XMM-Newton, an ESA science mission with instruments and contributions directly funded by ESA Member States and the USA (NASA). A.D. and G.M. acknowledge the CNES for financial support. F.B and M.M. acknowledges partial support from the ASI-INAF agreement n. I/009/10/0.
\end{acknowledgements}

\bibliographystyle{aa.bst}
\bibliography{paper.bbl}

\end{document}